\documentclass[twocolumn,showpacs,preprintnumbers,amsmath,amssymb]{revtex4}
\usepackage{graphicx}
\usepackage{dcolumn}
\usepackage{bm}

\newcommand{\ie}{{i.e.}}

\newcommand{\tr}{{\rm tr}}
\newcommand{\half}{\frac{1}{2} }

\def\ave#1{\langle #1 \rangle}

\def\deriv#1{\frac{\partial}{\partial#1}}
\def\dev#1#2{\frac{\partial#1}{\partial#2}}

\newcommand{\bfa}{{\cal A}\!\!\!\!\!{\cal A}}

\newcommand{\bfpi}{\large \pi\!\!\!\!\pi}
\newcommand{\jj}{{\cal J}\!\!\!\!\!{\cal J}}

\begin{document}

\title{Generalized Kubo formula for spin transport: A theory of linear response to non-Abelian fields}

\author{Pei-Qing Jin and You-Quan Li}

\affiliation{Zhejiang Institute of Modern Physics and Department
of Physics, Zhejiang University, Hangzhou 310027, People's
Republic of China }

\begin{abstract}

The traditional Kubo formula is generalized to describe the linear
response with respect to non-Abelian fields. To fulfill the demand
for studying spin transport, the SU(2) Kubo formula is derived by
two conventional approaches with different gauge fixings. Those
two approaches are shown to be equivalent where the
nonconservation of the SU(2) current plays an essential role in
guaranteeing the consistency. Some concrete examples relating spin
Hall effect are considered. The dc spin conductivity in response
to an SU(2) electric field vanishes in the system with parabolic
unperturbed dispersion relation. By applying a time-dependent
Rashba field, the spin conductivity can be measured directly. Our
formula is also applied to the high-dimensional representation for
the interests of some important models, such as Luttinger model
and bilayer spin Hall system.
\end{abstract}

\pacs{72.25.-b, 72.10.-d, 03.65.-w}

\received{\today}
\maketitle

\section{Introduction}

Kubo formula, one of the most important formulas in the linear
response theory, has been widely used in condensed matter physics
since it was derived by Kubo~\cite{Kubo} for the electrical
conductivity in solids. There are several kinds of Kubo formulas
for the external fields to which the system responses are
different. However, these formulas, such as those for the
electrical conductivity and the susceptibility, all describe
linear responses to the U(1) external fields.

Recently, a newly emerging field, spintronics~\cite{Wolf, Zutic},
has absorbed much attention for its promising applications in
quantum information storage and processing. Spin Hall
effect~\cite{Hirsch,Zhang,Niu0403,Kato,Wunderlich,Rashba,Hu,Schliemann,Niu0407,Sinitsyn,Shen,Inoue,Halperin,Dresselhaus},
as a candidate method to injecting spin current into
semiconductors, is also discussed intensively. In this effect, the
spin-orbit coupling is necessary no matter intrinsically or
extrinsically. As this coupling can be regarded as a contribution
of the SU(2) gauge potential~\cite{Li}, a new version of linear
response theory in SU(2) formulation is necessary. Nevertheless,
most of the previous works have mainly focused on the linear
response of such system to an external electric field, and hence
the traditional Kubo formula was adopted directly except
Ref~\cite{Schmeltzer} which dealt with non-Abelian response and
considered spin Hall effect in the presence of an SU(2) gauge
field. There were some papers~\cite{Nitta,Wang} discussing the
responses to a spin-orbit coupling with spatially varying
strength, but the authors employed other approaches rather than
Kubo formula as the SU(2) Kubo formula has not been established.
It thus becomes inevitable to develop a generalized Kubo formula
so that the linear response to the external SU(2) gauge fields can
be evaluated.

In present paper, we derive a formula which describes the linear
response to an SU(2) external field using the strategy ever
employed by Kubo for the U(1) case. It is not a straightforward
derivation since the algebra is totally different. Especially, the
expression of the SU(2) ``electric field '' evolves one more term
of gauge potentials~\cite{Yang} than the U(1) case due to its
non-Abelian feature. It seems obscure to directly find the
equivalence between the Kubo formulas derived with different gauge
fixings. We will show that the extra term in the SU(2) ``electric
field '' precisely corresponds to the nonvanishing term in the
``continuity-like'' equation~\cite{Li} which includes the spin
procession~\cite{Li,Niu0407}. Its origin stems from the definition
of the conserved current~\cite{Li} in the presence of the SU(2)
field. Since one of recent research interests focuses on the
spintronics, some explicit examples in spin Hall systems are
discussed in terms of our SU(2) Kubo formula, such as the spin
susceptibility and current in response to the effective spin-orbit
coupling~\cite{Rashba,Dresselhaus}. In spin Hall effect, the spin
conductivity is believed to be canceled by the effect of disorder
in two-dimensional electron gas~\cite{Inoue}. It is due to the
parabolic unperturbed dispersion relation~\cite{Das Sarma}. In
such a system, the spin current in response to the external
spin-orbit coupling also vanishes. Then the systems with
nonparabolic unperturbed dispersion relation become significant.
In such systems, the spin conductivity in response to either U(1)
or SU(2) external fields does not vanish. An experimentally
accessible case is also given in which the spin conductivity is
related to the dielectric function. We also extend the application
of our formula to a high-dimensional representation, saying
spin-$3/2$ representation, which is related to some important
systems, such as the Luttinger model~\cite{Luttinger} and the
bilayer systems~\cite{Boebinger}. The spin conductivity in the
Luttinger model vanishes, which describes the response to the
effective field of structural inversion asymmetry.

The paper is organized as follows. In
Sec.~\ref{sec:zerotemperature}, we derive a general Kubo formula
with respect to a single-frequency SU(2) external field at zero
temperature. Then we show in Sec.~\ref{sec:zerofrequency} that
this formulation is consistent with the one by choosing a
zero-frequency external field at the very beginning. In
Sec.~\ref{sec:onehalf}, we give the applications of our SU(2) Kubo
formula to some models of spin-$1/2$ representation. In
Sec.~\ref{sec:threehalf}, our theory is applied to a
high-dimensional representation (\ie, spin-$3/2$ representation)
system. Several concrete example models are also given. In
Sec.~\ref{sec:sum}, we give a brief summary with some remarks. In
the appendixes, we give the detailed calculations of the
correlation functions in Matsubara formalism.

\section{SU(2) Kubo formula at zero temperature}\label{sec:zerotemperature}

A generalized total Hamiltonian for a variety of models to study
spin transport can be written as~\cite{Li}
\begin{equation}\label{eq:totalHamiltonian}
H=\frac{1}{2m}\sum_l\Bigl(\mathbf{p}_l -\frac{e}{c}\mathbf{A}(\mathbf{r}_l,t)
  - \eta\bfa^a(\mathbf{r}_l,t)\tau^a \Bigr)^2 \!\! + \hat{V}_{dis},
\end{equation}
where $\hat V_{dis}$ is the potential caused by disorders.
Throughout this paper, the index $l$ refers to the $l$th particle,
$a$, $b$, and $c$ refer to spin space while $i$ and $j$ the
spatial space, and repeated indices are summed over. $\tau^a$
stands for the generators of SU(2) algebra, $\mathbf{A}$ and
$\bfa^a$ are the U(1) and SU(2) gauge potentials, respectively.
Usually these gauge potentials consist of two parts, internal and
external fields. In order to derive a general formula for the
conductivity in response to an SU(2) external ``electric field '',
we separate the Hamiltonian (\ref{eq:totalHamiltonian}) into two
parts, $H = H_0+H'$, with
\begin{eqnarray}
&& H_0= \frac{1}{2m} \sum_l \hat\bfpi_l^2 +\hat{V}_{dis}, \nonumber \\
&& H' = \frac{-\eta}{2m}\!\sum_l \Bigl(\hat\bfpi_l\cdot\bfa^a_{\rm ext}(\mathbf{r} _l,t)\tau^a
  +\tau^a\bfa^a_{\rm ext}(\mathbf{r}_l,t)\cdot\hat\bfpi_l\Bigr), \nonumber\\
\end{eqnarray}
where the operator $\hat\bfpi_l=\mathbf{p}_l -\frac{e}{c}\mathbf{A}_{\rm int}(\mathbf{r}_l,t)
  - \eta\bfa^a_{\rm int}(\mathbf{r}_l,t)\tau^a $
stands for the dynamical momentum involving internal U(1) and SU(2) potentials if any. Note that this separation
is up to the first order of $\bfa^a_{\rm ext}(\mathbf{r}_l,t)$.

The relation between SU(2) ``electric fields '' and the gauge
potentials is given by
\begin{equation}
E_i ^a =-\partial_0\mathcal{A}_i^a - \partial_i\mathcal{A}_0^a
+\eta\epsilon^{abc}\mathcal{A}_0^b\mathcal{A}_i^c,
\end{equation}
where $\eta$ is the coupling constant. The SU(2) ``electric field
''~\cite{Li} is expected to be realized by spatially~\cite{Nitta}
or timely~\cite{Wang} dependent Rashba or Dresselhaus coupling
strength. We consider the linear response with respect to the
``electric'' components of a non-Abelian field, $\mathbb{E}_i =
E^a_i\tau^a$. For simplicity, we take the SU(2) external field to
be of single-frequency, namely,
\begin{equation}\label{single-frequency elecfield}
E^a_i (\mathbf{r},t)= E^a_i (\mathbf{q},\omega)~
 e^{i\mathbf{q}\cdot\mathbf{r}-i\omega t}.
\end{equation}
The problem involving the external field of an arbitrary form on
$\mathbf{r} $ only differs from this case by a fourier transform.
As is well-known, the external field of frequency($\mathbf
q,\omega$) actually reads $E^a_i(\mathbf q,\omega)\cos(\mathbf
q\cdot \mathbf r-\omega t)$ which is real valued. The expression
in Eq.~(\ref{single-frequency elecfield}) is convenient for
calculation. Thus in the subsequent formulas, only the real parts
have the physical meaning.

Now we choose a gauge that $\mathcal{A}_0^a=0$, which corresponds
to the temporal gauge in the U(1) case, then the SU(2) external
``electric field'' comes into the Hamiltonian through the vector
potential $\mathcal{A}_i^a$,
\begin{equation}
\mathcal A_i^a(\mathbf r,t) = \frac{1}{i \omega} E_i ^a(\mathbf r,t)=\frac{1}{i \omega} E_i ^a(\mathbf
q,\omega)e^{i\mathbf q\cdot\mathbf r-i\omega t}.
\end{equation}
Hereafter, we omit the subscript specifying the external field for simplicity.

Based on the definition for single particle, one can define the
SU(2)-current operator for the unperturbed system
\begin{equation}\label{SU(2)current}
\hat{\mathbf J}^a  (\mathbf{r})
  = \frac{\eta}{4m} \sum_l\Bigl[\{\tau^a, \hat\bfpi_l \}
   \delta(\mathbf{r}-\mathbf{r}_l)
    +\delta(\mathbf{r}-\mathbf{r}_l)\{\tau^a, \hat\bfpi_l \}\Bigr],
\end{equation}
where the curl bracket denotes anticommutator and the velocity
operator of the $l$th particle determined by the Heisenberg
equation of motion, $\hat\bfpi_l/m=[\mathbf{r}_l,~H_0]/(i\hbar)$
is matrix-valued for the SU(2) case. In terms of this current, the
perturbation Hamiltonian $H'$ can be expressed as
\begin{eqnarray}
H' &=& -\int d\mathbf r~\hat{J}_i ^a (\mathbf{r})\mathcal{A}_i ^a (\mathbf{r} ,t) \nonumber \\
   &=& -\frac{1}{i \omega} \  \hat{J}_i ^a (\mathbf{q}) \  E_i ^a (\mathbf{q} ,\omega) e^{-i \omega
   t},
\end{eqnarray}
where $\hat{J}_i ^a (\mathbf{q})$ is the Fourier image of
$\hat{J}_i^a (\mathbf{r})$.
Clearly, the interaction term is the
product of the SU(2) current and the SU(2) ``electric field ''.

Taking the perturbation of the external field into account, we
have $\mathbf{\Pi}_l=\hat \bfpi_l -
\eta\bfa^a(\mathbf{r}_l)\tau^a$. Then the total SU(2) current
driven by the external SU(2) ``electric field '' reads
\begin{eqnarray}\label{eq:totoalcurrent}
\jj^a\! (\mathbf{r}, t) \!\!\!&=&\!\!\! \frac{\eta}{4m}
 \sum_l \Bigl[
  \{\tau^a ,\mathbf{\Pi}_l \}\delta(\mathbf{r}-\mathbf{r}_l)
   +\delta(\mathbf{r}-\mathbf{r}_l)\{\tau^a, \mathbf{\Pi}_l \} \Bigr]
  \nonumber \\
 &=& \!\!\hat{\mathbf J}^a (\mathbf{r}) - \frac{\eta^2}{4m}\bfa^a (\mathbf{r}, t)~n_0,
\end{eqnarray}
where $n_0$ is the particle density.

At zero temperature, this SU(2) current is evaluated for the
ground state of the system. In the interaction representation, the
state $|\psi(t)\rangle$ of the system at time $t$ is related to
the eigenvector $|\phi\rangle$ of $H_0$ by the S-matrix, \ie,
$|\psi(t)\rangle = S(t,-\infty)|\phi\rangle$. Up to the linear
order of $H'_I(t')$,
$S(t,-\infty)=1-\frac{i}{\hbar}\int^t_{-\infty} dt' H'_I (t')$,
where $H'_I (t')= e^{i H_0 t'/\hbar}H' e^{-i H_0 t'/\hbar}$. Then
the average of the first term in the total SU(2) current
Eq.~(\ref{eq:totoalcurrent}) is given by
\begin{eqnarray}
\ave{\hat{J}_i ^a (\mathbf{r},t)}
 =\ave{\hat{J}_i ^a(\mathbf{r},t)}_0
  - \!\frac{i}{\hbar}\int_{-\infty}^{~t}\!\!dt'
  \ave{[\hat{J}_i^a (\mathbf{r},t),H'_I(t')\ ]}_0
  \nonumber \\
= \frac{ E^b_j(\mathbf{r},t)}{\hbar\omega}\!
   \int_{-\infty}^{~t}\!\!dt'
 e^{i\omega(t-t')} e^{-i\mathbf{q}\cdot\mathbf{r}}
  \ave{[\ \hat{J}_i^a (\mathbf{r}, t),
   \hat{J}_j^b (\mathbf{q},t')]}_0, \nonumber\\
\end{eqnarray}
where $\ave{\hat{J}_i ^a (\mathbf{r} ,t)}_0$ has been dropped
since no SU(2) current is considered to follow in the absence of
the external fields. Together with the second term, we obtain the
following expression:
\begin{eqnarray}
&&\ave{\mathcal J_i^a (\mathbf{r}, t)}
  =\ave{\hat{J}_i^a (\mathbf{r}, t)}
   -\frac{\eta^2}{4m}\mathcal{A}_i^a (\mathbf{r}, t)~n_0
       \nonumber \\
&&=\frac{E^b_j(\mathbf{r}, t)}{\hbar\omega}
\int_{-\infty}^{~t}\!\!dt'
 e^{i\omega (t-t')} e^{-i \mathbf{q}\cdot\mathbf{r}}
  \ave{[\hat{J}_i^a(\mathbf{r}, t),\hat{J}_j^b(\mathbf{q}, t')] }_0
   \nonumber\\
&& \hspace{6mm} +\frac{i\eta^2 n_0}{4m\omega} \delta^{ab}\delta_{ij} E^b_j(\mathbf{r}, t)
     \nonumber \\
&&   \equiv \sigma^{ab}_{ij}(\mathbf q,\omega; \mathbf{r})E^b_j(\mathbf r,t).
\end{eqnarray}
Since the conductivity represents the property of the whole system,
we need take the average over the system to get
the SU(2) conductivity, 
\begin{eqnarray}\label{SU(2)conductivity}
\sigma _{ij} ^{a b} (\mathbf{q},\omega)&=&
 \frac{1}{\hbar\omega V} \int _{-\infty}^{~t}\!\!dt' \  e^{i \omega (t-t')}
  \ave{[\hat{J}_i^{\dagger a} (\mathbf{q} ,t),
    \hat{J}_j^b (\mathbf{q}, t') ] }_0
     \nonumber\\
   &&  +\frac{i\eta^2~n_0}{4m\omega} \delta^{ab}\delta_{ij},
\end{eqnarray}
with $V$ the volume of the system.
The spin conductivity here is a tensor in spin space
rather than a vector as in the case of
linear response to the U(1) external field.

As a conventional strategy, a retarded current-current correlation
function is thus introduced to calculate this conductivity,
\begin{equation}\label{eq:correlation}
Q_{ij}^{ab} (\mathbf{q}, t-t')=
 -\frac{i}{V}~\theta(t-t')
  \ave{[\ \hat{J}^{\dag a}_i (\mathbf{q},t),\
    \hat{J}_j^b (\mathbf{q}, t')~] }_0,
\end{equation}
where $\theta(t-t')$ is the step function which vanishes unless
$t>t'$. The Fourier transform of Eq.~(\ref{eq:correlation}) is
given by
\begin{eqnarray}
&& Q_{ij}^{ab} (\mathbf{q} ,\omega)
    \nonumber \\
&& =-\frac{i}{V}\!\!\int_{-\infty}^{+\infty}\!\!\!\! dt ~\theta(t-t') e^{i \omega (t-t')}
      \ave{[ \hat{J}^{\dag a}_i (\mathbf{q} ,t),~ \hat{J}_j ^b (\mathbf{q} ,t')\ ]}_0.
    \nonumber \\
\end{eqnarray}
Comparing with Eq.~(\ref{SU(2)conductivity}), we obtain
\begin{equation}\label{SU(2)Kubo formula}
\sigma _{ij} ^{a b} (\mathbf{q} ,\omega)
 =\frac{i}{\hbar\omega} \Bigl[ Q_{ij}^{ab} (\mathbf{q} ,\omega)
 +\frac{\hbar\eta^2 n_0}{4m} \delta^{ab} \delta_{ij} \Bigr].
\end{equation}
To simplify the calculations, we introduce a Matsubara function
$Q_{ij}^{ab} (\mathbf{q}, i\nu)$ which reduces to the retarded
correlation function $Q_{ij}^{ab} (\mathbf{q}, \omega)$ by
changing $i\nu$ to $\omega+i\delta$,
\begin{equation}
Q_{ij}^{ab} (\mathbf{q} , i\nu)=-\frac{1}{V} \int_{0}^{\beta} du \  e^{i \nu u} \ave{T_u \ \hat{J}^{\dag a}_i
(\mathbf{q} ,u) \hat{J}_j ^b (\mathbf{q} ,0)},
\end{equation}
where $T_u$ denotes the $u$-ordering operator and
$\beta=(k_B^{\,}T)^{-1}$ with $k_B^{\,}$ the Boltzmann constant. We thus
have derived a generalized Kubo formula for spin transport in
response to an external SU(2) ``electric field ''.

\section{An equivalent formulation for zero frequency}\label{sec:zerofrequency}

In the previous section, we derived the SU(2) Kubo formula
choosing the gauge potential $\mathcal{A}_0^a=0$. To obtain the dc
conductivity, one just needs to take the limit $\omega\rightarrow
0$. As is well-known in the conventional electrical conductivity,
the Kubo formula can also be derived alternately by choosing a
constant external field as a start point. The continuity equation
for electric charge conservation guarantees the two derivations to
be equivalent. Whereas, in the SU(2) case, the current defined by
Eq.~(\ref{SU(2)current}) is not conserved as long as an SU(2)
interaction is present. For example, the spin current, a special
SU(2) current with $\eta=\hbar$, is not conserved if there exists
the Zeeman term or spin-orbit coupling. In this case, the
continuity equation does not hold~\cite{Li,Schmeltzer}, instead,
we have the following relation:
\begin{equation}\label{continuity-like-equation}
 \bigl(\deriv{t}-\eta\vec{\mathcal A}_0\times\bigr)\vec{\sigma}(\mathbf{r},~t)
 +\big(\deriv{x_i}+\eta\vec{\mathcal A}_i\times\bigr)
  \vec{J}_i(\mathbf{r},~t) =0,
\end{equation}
where $\sigma^a(\mathbf{r})=\eta\psi^\dag(\mathbf{r})\tau^a
\psi(\mathbf{r} )$ and $\mathbf{J}^a(\mathbf{r}, t)$ are the SU(2)
density and current respectively, and notations
$\vec\sigma=(\sigma^1,~\sigma^2,~\sigma^3)$, $\vec{\mathcal
A}=({\mathcal A}^1,~{\mathcal A}^2,~{\mathcal A}^3)$ etc. are
adopted. Unlike the charge current which is conserved, the spin
current is not conserved, thus a natural question is whether the
SU(2) Kubo formula we derived in the previous section is still
consistent with the other derivation?

Now let us choose $\partial_0 \mathcal A^a_i=0$ for the zero frequency case, then the SU(2) electric field and
the perturbation Hamiltonian are given by
\begin{equation}\label{SU(2) electric field}
E^a_i=-\partial_i\mathcal{A}^a_0
 +\eta\epsilon^{abc}\mathcal{A}^b_0\mathcal{A}^c_i,
\end{equation}
and
\begin{equation}
H'=\int d\mathbf r~\sigma^a(\mathbf{r}, t)\mathcal{A}^a_0 (\mathbf{r}).
\end{equation}

By means of the method suggested by Luttinger,
the total SU(2) current can be obtained once the density matrix $\rho$
is introduced.
The density matrix including the deviations caused by the perturbation
takes the form
\begin{equation}
\rho(t)=\rho_0^{~}+\delta\rho(t),
\end{equation}
where $\rho_0^{~}$ refers to the density matrix with respect to
the unperturbed Hamiltonian and $\delta\rho(t)$ is brought about
by the perturbation one, $H'$. From the equation of motion for the
perturbed part of the density matrix,
\begin{equation}
i\hbar\dev{\delta\rho(t)}{t} =[H_0,\delta\rho(t)]+[H'\!,\,\rho_0^{~}],
\end{equation}
we can obtain a solution for $\delta\rho(t)$
\begin{equation}
\delta\rho(t)=-\frac{1}{\hbar}\int^{\infty}_0 dt \int_0^\beta
d\beta' ~\rho_0^{~} \deriv{t} H'_I(-t-i\beta').
\end{equation}
With the help of the density matrix, the SU(2) current can be then
evaluated by taking the average
\begin{eqnarray}\label{SU(2)current density matrix}
\ave{\hat{J}_i ^a (\mathbf{r} ,t)} =
      \tr \bigl(\rho(t) \hat{J}_i^a (\mathbf{r}) \bigr) \hspace{43mm}
  \nonumber \\
\!\! =-\frac{1}{\hbar}\int^{\infty}_0\!\!\!dt \int_0^\beta
 \!d\beta'\tr\Bigl[\rho_0^{~} \ \deriv{t} H'_I(-t-i\beta') \ \hat{J}_i^a
     (\mathbf{r}) \Bigr], \hspace{1mm}
    \nonumber\\
\end{eqnarray}
where the equilibrium part of the current
$\tr(\rho_0^{~}\hat{J}_i^a (\mathbf{r} ))$ is assumed to be zero.
The derivative of $H'_I$ with respect to time $t$ is calculated as
\begin{equation}
\partial_t H'_I(-t)=\int d\mathbf r~\partial_t \sigma^a(\mathbf{r},-t)\mathcal{A}^a_0(\mathbf{r}).
\end{equation}
Using the ``continuity-like'' equation
(\ref{continuity-like-equation}) and integration by parts, we have
\begin{eqnarray}
\partial_t H'_I(-t)&=&\int d{\mathbf r}
  \Bigl(\eta\epsilon^{abc}(\mathcal{A}_0^b\,\sigma^c-\mathcal{A}_i^b J^c_i)-\partial_i J_i^a\Bigr)
   \mathcal{A}^a_0
     \nonumber \\
  &=& \! -\int d\mathbf r ~E_i^a J^a_i(\mathbf{r},-t),
\end{eqnarray}
where we did not write out the arguments in the first line for simplicity.
Substituting it into Eq.~(\ref{SU(2)current
density matrix}), we obtain
\begin{eqnarray}
&&\ave{\hat{J}_i^a (\mathbf{r},t)}
  \nonumber\\
 && =\!\frac{1}{\hbar}\int^{\infty}_0\!\!\!\! dt\!\int_0^\beta\!\!\! d\beta'\!\!\int\!\! d\mathbf{r}'
   ~\tr \Bigl[\rho_0^{~}
    E_j^b J^b_j(\mathbf{r}',-t-i\beta')\hat{J}_i^a (\mathbf{r} ) \Bigr].
     \nonumber \\
\end{eqnarray}
Consequently, the dc SU(2) conductivity is obtained
from the above equation after integrating $\mathbf{r}$ over the volume $V$,
\begin{equation}
\sigma_{ij}^{ab}= \frac{1}{\hbar V}\int^{\infty}_0 dt \int_0^\beta
d\beta' \ \tr \Bigl[\rho_0^{~} \ J^b_j(-t-i\beta') \ \hat{J}_i ^a \Bigr].
\end{equation}

This result is obviously independent on the frequency. It is also
consistent with the one which we derived in the previous section
once we introduce the representation of the eigenstates
$|n\rangle$ of $H_0$. Note that the spin procession terms,
$\eta\vec{\mathcal A}_i\times
\vec{J}_i(\mathbf{r},~t)-\eta\vec{\mathcal
A}_0\times\vec{\sigma}(\mathbf{r},~t)$, precisely compensate the
second term of Eq.~(\ref{SU(2) electric field}), which makes our
theory self-consistent. Since the SU(2) ``electric field ''
includes an extra term of gauge potential in comparison to the
U(1) field, the nonconservation of the SU(2) current plays an
essential role in guaranteeing the consistency. That is to say,
the SU(2) current exactly responds to the SU(2) ``electric field
'' no matter which gauge is chosen.

\section{Applications for spin-$1/2$ representation}\label{sec:onehalf}

From now on, we will give some applications of our SU(2) Kubo formula. In this section, we mainly focus on the
examples in spin-$1/2$ representation without impurities and in the limit $q\rightarrow 0$.

\subsection{Spin susceptibility}

Spin is a category of SU(2) entity with $\eta=\hbar$. The spin
degree of freedom is discussed extensively in recent years for its
promising application. The effective spin-orbit coupling emerged
significantly in some semiconductors~\cite{Rashba,Dresselhaus} is
of importance for its possible manipulating of spin. Using our
SU(2) Kubo formula, we can directly calculate the spin
susceptibility which describes the linear response of the spin
density to the spin-orbit coupling.

The spin susceptibility $\chi^{ab}_i$ is defined as
\begin{equation}
\ave{\hat S^a}=\chi^{ab}_i E^b_i,
\end{equation}
where $\hat S^a=\hbar\sum_k C^\dag_k \tau^a C_k$ is the spin
density. Here we adopted a simplified notion
$C^\dag_k=(C^\dag_{k\uparrow},~C^\dag_{k\downarrow})$ with
$C^\dag_{k\uparrow}$ creating a spin-up particle of momentum $k$
etc. The corresponding retarded correlation function in Matsubara
formalism $\Pi^{ab}_i (i\nu)$ is given by
\begin{equation}
\Pi_i^{ab} (i\nu)=-\frac{1}{V} \int_{0}^{\beta} du \  e^{i \nu u} \ave{\rm T_u \, \hat{S}^a (u) \hat{J}_i ^b
(0)}.
\end{equation}
Hereafter, we take the unperturbed Hamiltonian to be $H_0=\sum_k
C^\dag_k(\varepsilon(k)+d^a(k)\tau^a) C_k$ for its elegant form in
Green's function. The second term represents the internal SU(2)
field with $d^a$ the components of this field. This system has two
bands, $E_{-}=\varepsilon(k) + |d|$ and
$E_{+}=\varepsilon(k)-|d|$, with $|d|=\sqrt{d^a\,d^a}$. In the
limit $\omega\rightarrow 0$, the susceptibility reads
\begin{equation}
\chi_i^{ab} =\frac{\hbar}{2V} \sum_k
  \frac{n_{F_-}-n_{F_+}}{|d|^3}~\epsilon^{abc} d^c
   \dev{\varepsilon(k)}{k_i},
\end{equation}
where $n_{F_-}$ and $n_{F_+}$ are the Fermi distribution functions
and ``$-$, $+$" label the different bands. This result is
antisymmetric to the indices labeling spin degree of freedom.
Using this result, we calculate the spin susceptibilities with two
kinds of internal fields, Rashba and Dresselhaus couplings,
\begin{eqnarray}
H_{_R} &=& -2\alpha (k_x\tau^y-k_y\tau^x), \nonumber\\
H_{_D} &=& -2\beta (k_x\tau^x-k_y\tau^y).
\end{eqnarray}
These two kinds of couplings dominate in narrow gap semiconductors
such as GaAs and here we take their two-dimensional (2D) forms to
represent the effective spin-orbit couplings in two-dimensional
electron gas (2DEG). In these cases, the components $\chi_i^{xy}$
vanish since $d^z=0$.
The results are shown in Table I, where we have taken the usual
parabolic form that $\varepsilon(k)=\hbar^2 k^2/2m$.
\begin{table}
\renewcommand{\arraystretch}{1.5}
\begin{tabular}{| c | c | c | c | c |}
  \hline
  \hspace{2mm} SU(2) internal field\hspace{2mm} & \ $\displaystyle\chi_{x}^{zx}$ \
  & \ $\chi_{x}^{zy}$ \ & \ $\chi_{y}^{zx}$ \
  & \ $\chi_{y}^{zy}$ \\[1mm]
  \hline
  Rashba ($\frac{\hbar}{32\pi\alpha}$)
  & 1 & 0 & 0 & 1 \\[1mm]
 \hline
  Dresselhaus ($\frac{\hbar}{32\pi\beta}$)
  & 0 & -1 & -1 & 0 \\[1mm]
  \hline
\end{tabular}
\caption{ Spin susceptibilities:  $\alpha$ and $\beta$ are coupling constants for the Rashba and
Dresselhaus coupling respectively.}
\end{table}

\subsection{Spin conductivity}

With $\eta=\hbar$, the spin current reads
\begin{equation}
\hat{J}_i ^a=\frac{1}{2}\sum_k C_k^\dag \Bigl\{\dev{\varepsilon(k)}{k_i}+\dev{d^b}{k_i} \tau^b,\tau^a \Bigr\}
C_k.
\end{equation}
After calculating the Matsubara function (see Appendix A) and
changing $i\nu\rightarrow\omega$, we derive the conductivity
\begin{equation}\label{eq:spin conductivity}
\sigma_{ij}^{ab} =\frac{1}{2V}\sum_k \frac{n_{F-}-n_{F+}}{|d|^3}\,
 \epsilon^{abc} d^c \dev{\varepsilon(k)}{k_i}
\dev{\varepsilon(k)}{k_j}.
\end{equation}
This expression manifests that the conductivity is antisymmetric
to the spin indices $a, b$ and symmetric with respect to the
spatial indices $i, j$ with the parabolic dispersion relation.
Note that when the U(1) part of $H_0$ is parabolic, \ie,~
$\varepsilon(k)=\hbar^2 k^2/2m$, and $d^a$ is linear of $k^a$, the
summation over $k$ vanishes. Since the conventional spin-orbit
couplings are Rashba and Dresselhaus couplings, which contain no
quadratic terms of $k$, we should consider
$\varepsilon(k)=\frac{\hbar^2}{2m} \bigl[ (k_x+c)^2 +(k_y+c')^2
\bigr]$ for nonvanishing results, which represents a shift of
momentum $k$ in the material. Table II shows the spin
conductivities with two kinds of internal fields.
\begin{table}
\renewcommand{\arraystretch}{1.5}
\begin{tabular}{|c|c|c|c|c|c|c|}
  \hline
 SU(2) internal field& \ $\sigma_{xx}^{zx}$ \ & \ $\sigma_{xx}^{zy}$ \ & \ $\sigma_{xy}^{zx}$ \
  & \ $\sigma_{xy}^{zy}$ \ & \ $\sigma_{yy}^{zx}$ \ & \ $\sigma_{yy}^{zy}$ \\
  \hline
  Rashba  ($\frac{\hbar^2}{16\pi m \alpha}$)
  & c & 0 & $\frac{c'}{2}$ & $\frac{c}{2}$ & 0 & $c'$ \\
 \hline
  \hspace{1mm}Dresselhaus ($\frac{\hbar^2}{16\pi m
  \beta}$)\hspace{1mm}
  & 0 & -c & $-\frac{c}{2}$ & $-\frac{c'}{2}$ & $-c'$ & 0 \\
 \hline
\end{tabular}
\caption{Spin conductivities: $c$ and $c'$ represent the displacements in $k$ space.}
\end{table}

At this stage, it is worthwhile to recall some previous work in
spin Hall effect. Up to now, a general consensus is made that the
spin conductivity in response to an external Maxwell electric
field is exactly canceled by the effect of disorder in
two-dimensional electron gas with spin-orbit coupling. The
cancellation is due to the parabolic form of unperturbed band
structure~\cite{Das Sarma}. It is worthwhile to point out that our
SU(2) conductivity also vanishes when $\varepsilon(k)$ takes the
parabolic form even in the absence of disorder. It is an essential
difference that our conductivity refers to the linear response to
an external Yang-Mills electric field which is also a vector in
SU(2) Lie algebra space whose bases, the Pauli matrices, are
anticommute. Anyway the system with nonparabolic dispersion
relation is of great importance, since the conductivity, no matter
in the usual spin Hall effect with disorder or derived by our
SU(2) Kubo formula without disorder, is expected to be observed in
experiments.

Finally, we will consider an experimentally available case. Since the Rashba coupling strength can be tuned by
the gate voltage applied to 2DEG, we take a Rashba coupling with time-dependent strength as the external SU(2)
field and Dresselhaus coupling as an internal field. Then we can get an ac conductivity depending on the
frequency $\omega$. The result reads
\begin{equation}
\sigma^{zy}_{xx}(\omega)=-\frac{c\hbar^6\omega^2}{32m^2\beta^3\pi}
\epsilon_{_D}(\omega),
\end{equation}
where $\epsilon_{_D}(\omega)$ is the dielectric function
caused by the Dresselhaus spin-orbit
coupling~\cite{Rashba04}, namely,
\begin{equation}
\epsilon_{_D}(\omega)=\frac{4\beta^3}{\hbar^2\omega^2} \int^{k_{F-}}_{k_{F+}}\frac{k^2dk}{(2\beta k)^2-\hbar^2\omega^2}.
\end{equation}
This dielectric function is a macroscopic quantity and can be directly measured.
Carrying out the
integration over $k$ gives a resonant result,
\begin{equation}
\sigma^{zy}_{xx}=-\frac{c\hbar^2}{16\pi\beta
m}-\frac{c\hbar^4\omega}{128\pi\beta^3 m^2}\ln
\Bigl|\frac{k-\hbar\omega/2\beta}{k+\hbar\omega/2\beta}\Bigr|^{k_F-}_{k_F+},
\end{equation}
where $k_F-$ and $k_F+$ refer to the Fermi momenta of both bands.
The same resonance is also shown in Ref.~\cite{Bryksin}. Other
components are given by $\sigma^{zx}_{xy}=\half\sigma^{zy}_{xy}$
while $\sigma^{zx}_{yy}$ and $\sigma^{zy}_{xy}$ differ from them
by $c\rightarrow c'$.

\section{Applications for spin-$3/2$ representation}\label{sec:threehalf}

In the previous section, we have discussed several examples using
the SU(2) Kubo formula in spin-$1/2$ representation. It is well-
known that there exit many important systems which carry out the
spin-$3/2$ representation of the SU(2) algebra, for example, the
Luttinger model~\cite{Luttinger} containing the intrinsic
spin-orbit coupling, bilayer systems~\cite{Boebinger} taking into
account of spin degree of freedom, etc. Thus it is worthwhile for
us to extend our discussion to high-dimensional representations,
such as spin-$3/2$ representation. The examples mentioned above
are also discussed, which may be instructive for the experiments.

\subsection{General consideration}

The spin-$3/2$ representation of   SU(2) generators read
\begin{eqnarray}
\tau^x &=&\left(
    \begin{array}{cccc}
                0 & {\sqrt 3}/2 &       0  & 0 \\
      {\sqrt 3}/2 &          0  &       1  & 0 \\
                0 &          1 &        0  & {\sqrt 3}/2 \\
                0 &          0 & {\sqrt 3}/2 & 0 \\
    \end{array}
    \right),
    \nonumber\\[2mm]
\tau^y &=& \left(
    \begin{array}{cccc}
    0 & -{i\sqrt 3}/2 &       0 & 0 \\
{i\sqrt 3}/2 &      0 &      -i & 0 \\
    0 &             i &       0 & -i\sqrt 3/2 \\
    0 &             0 &  i\sqrt 3/2 & 0 \\
    \end{array}
    \right),
    \nonumber\\[2mm]
\tau^z &=&\left(
    \begin{array}{cccc}
      3/2 ~&    0 &    0 & 0 \\
        0 ~&  1/2 &    0 & 0 \\
        0 ~&    0 & -1/2 & 0 \\
        0 ~&    0 &    0 & -3/2 \\
    \end{array}
    \right).
\end{eqnarray}
To simplify the calculations, we use a convenient representation
of the Clifford algebra adopted in Ref.~\cite{Zhang04}, namely,
$\Gamma^1 = \sigma^z \otimes \sigma^y $, $\Gamma^2 = \sigma^z
\otimes \sigma^x $, $\Gamma^3 = \sigma^y \otimes I $, $\Gamma^4 =
\sigma^x \otimes I$ and $\Gamma^5 = \sigma^z \otimes \sigma^z$
where the $\sigma$'s are the 2 by 2 Pauli matrices. These gamma
matrices satisfy $\{\Gamma^\alpha,\Gamma^\beta
\}=2\delta_{\alpha\beta}$ and $\Gamma^1 \Gamma^2 \Gamma^3 \Gamma^4
\Gamma^5=-1$. Hereafter, the Greek indices $\alpha, \beta,
\gamma,$ etc. run from 1 to 5. This representation can be obtained
from the Dirac representation of the gamma matrices by a unitary
transformation. Those five matrices can compose ten antisymmetric
matrices $\Gamma^{\alpha\beta}=[\Gamma^\alpha,\Gamma^\beta]/2i$
which constitute the spinor representation of SO(5) algebra.  The
spin-3/2 operators can be expressed as a linear combination of
those $\Gamma^{\alpha\beta}$, \ie, \
$\tau^a=\frac{1}{4i}L^a_{\alpha\beta}\Gamma^{\alpha\beta}$, where
the coefficients $L^a_{\alpha\beta}=-L^a_{\alpha\beta}$ are just
the five-dimensional representation of the SU(2) algebra (\ie,
spin-2 representation of the angular momentum operators).

Since $\Gamma^\alpha$,  $\Gamma^{\alpha\beta}$ together with the identity $I$ span the space of $4\times4$
Hermitian matrices, one can write out a general Hamiltonian in spin-$3/2$ representation in  terms of those
gamma matrices,
\begin{equation}
H_0=\sum_k C^\dag_k(\varepsilon (k)+d^\alpha(k) \Gamma^\alpha)C_k,
\end{equation}
where $C^\dag_k=(C^\dag_{k, 1},\,C^\dag_{k,2},\, C^\dag_{k, 3},\,
C^\dag_{k,4})$ with the second index referring to either spin-band
or spin-layer labels. Here we do not include the linear
combination of $\Gamma^{\alpha\beta}$ which makes the Green's
functions difficult to calculate. For this unperturbated
Hamiltonian, there exist two types of perturbation part $H'$. One
is constructed by $\Gamma^{\alpha\beta}$ and the other by
$\Gamma^\alpha$. The problem relating the spin current in
Luttinger model in response to the effective field caused by the
structure inversion asymmetry is of the first type. In this case,
the structure inversion asymmetry is taken as the perturbation and
hence
\begin{equation}
H'=\sum_k C^\dag_k h^a \tau^a C_k.
\end{equation}
Then the linear response of the spin current to $H'$ reads
\begin{equation}
\ave{\hat J^a_i}=\sigma^{ab}_i h^b.
\end{equation}
In calculating the retarded correlation function $Q^{ab}_{i}(i\nu)$, we will encounter
\begin{widetext}
\begin{eqnarray}\label{tr}
&&\tr \bigl(G(k,i\omega_n) \hat{J}^a_i
  G(k,i\omega_n+i\nu) \tau^b \bigr)   \nonumber \\
 &=&-\frac{1}{16}\tr \Bigl( G(k,i\omega_n)\Bigl\{{\dev{\varepsilon(k)}{k_i}
  + \dev{d^\alpha}{k_i}\Gamma^\alpha,~
  L^a_{\beta\gamma}\Gamma^{\beta\gamma}}\Bigr\} G(k,i\omega_n+i\nu) \ L^b_{\mu\nu}\Gamma^{\mu\nu} \Bigr),
\end{eqnarray}
\end{widetext}
where $G(k,i\omega_n)$ is the Matsubara function for which the
detailed calculation is given in Appendix B.

Since the traces of gamma matrices are always real, the appearance
of double $\Gamma^{\alpha\beta}$ matrices makes Eq.~(\ref{tr})
real and the summation of the Matsubara function also gives no
imaginary contribution after changing $i\nu\rightarrow\omega$.
This directly results in a vanishing spin conductivity.

The second type of perturbation is constructed by $\Gamma^\alpha$,
\begin{equation}
H'=\sum_k C^\dag_k h^\alpha \Gamma^\alpha C_k,
\end{equation}
and later we will discuss some concrete examples. The SU(2) Kubo
formula is then
\begin{equation}
\ave{J^b_i}=\sigma^{b\alpha}_i h^\alpha.
\end{equation}
The corresponding retarded correlation function is given by
\begin{equation}
Q^{b\alpha}_i (i\nu)=
  -\frac{1}{V} \int^{\beta}_0 du~ e^{i\nu u}\ave{\rm T_u ~
  \hat{J}^{\dag b}_i (u) \hat\Gamma^\alpha (0) }_0.
\end{equation}
After changing $i\nu\rightarrow\omega$ and taking the limit
$\omega\rightarrow0$, we obtain the dc conductivity
\begin{widetext}
\begin{equation}
\sigma^{b\alpha}_i  = \frac{-\eta}{4\hbar V}\sum_k \textrm{Im}
 \Bigl(2\dev{\varepsilon(k)}{k_i}L^b_{\alpha\beta}d^\beta
  -\epsilon^{\alpha\beta\gamma\mu\nu}L^b_{\beta\gamma}
  \dev{d^\mu}{k_i}d^\nu \Bigr)\frac{n_{F_-}-n_{F_+}}{|d|^3}.
\end{equation}
\end{widetext}

\subsection{Concrete examples}

Now we are in the position to discuss two concrete examples with
the second type of $H'$. First, we calculate the spin conductivity
for a bilayer system undergoing the Rashba coupling along opposite
directions,
\begin{equation}
H_0=\varepsilon (k) +\alpha\sigma^z\otimes(k_x\sigma^y-k_y\sigma^x)+\xi\sigma^x\otimes I.
\end{equation}
In this model the tunneling between the two layers is included in
which $\xi$ accounts for the tunneling strength. We take the SU(2)
flux to be the external perturbation,
$H'=\phi~\sigma^z\otimes\sigma^z$, \ie, \ $h_5=\phi$. A direct
calculation of the spin conductivity gives the following result:
\begin{eqnarray}
\sigma^{z5}_i \!\! &=&\!\! 0,
    \nonumber \\
\sigma^{x5}_x \!\!&=&\!\! -\frac{1}{4\pi\alpha\sqrt{\alpha^2
                   k^2+\xi^2}} \Bigl(\frac{\sqrt 3}{4}
                  \frac{\hbar^2(\alpha^2 k^2
                  +2\xi^2)}{m\alpha^2}+\xi\Bigr)^{k_{F-}}_{k_{F+}},
      \nonumber \\
\sigma^{y5}_y \!\!&=&\!\! -\sigma^{x5}_x.
\end{eqnarray}
In the limit $\xi\rightarrow 0$, $\sigma^{x5}_x$ reduces to
$-\frac{\sqrt 3}{4\pi\alpha}$.

As another example, we take the tunnelling term to be the
perturbation, that is $H'=\xi'\sigma^x\otimes I$ and
$H_0=\varepsilon (k)
+\alpha\sigma^z\otimes(k_x\sigma^y-k_y\sigma^x)$, correspondingly,
$h^4=\xi'$. Then we have the following result:
\begin{eqnarray}
\sigma^{z4}_\alpha &=& 0, \nonumber \\
\sigma^{x4}_x &=& -\frac{1}{8\pi\alpha}, \nonumber \\
\sigma^{y4}_y &=& -\sigma^{x4}_x.
\end{eqnarray}

\section{Summary and remarks}\label{sec:sum}

In this paper, we have generalized the Kubo formula to describe
the linear response of the SU(2) current to the external SU(2)
``electric field", which traditionally describes the one to the
U(1) external field. From two distinct routes with different gauge
fixings, we derived the SU(2) Kubo formula and showed that those
two approaches are equivalent. The non-Abelian feature of SU(2)
electric field involves one more term of gauge potentials in
comparison to the U(1) case, while this term precisely compensates
the nonconservation part in the SU(2) continuity-like equation for
the SU(2) current.

For the interests in spin transport, we applied our formula to
calculate the spin susceptibility and spin conductivity in the
system containing a Rashba or Dresselhaus field. The results show
that in the usual system, where $\varepsilon(k)=\hbar^2 k^2/2m$,
the spin susceptibility is constant. However, the spin
conductivity vanishes, much like the case in the spin Hall effect
where the spin conductivity in response to the external electric
field vanishes in the presence of disorder. To derive the
nonvanishing spin conductivity, the systems with nonparabolic
unperturbed band structure are necessary, and the spin
conductivity, no matter in response to the U(1) or SU(2) electric
field, is expected to be observed in such systems. What is more,
we also discussed an experimentally available case. In response to
the time-dependent Rashba field, the spin conductivity is related
to the dielectric function which can be measured directly.

Generalized to the high-dimensional representation, our SU(2) Kubo
formula is available to discuss the Luttinger model as well as
bilayer spin Hall effect. The spin conductivity in response to the
effective field caused by structural inversion asymmetry in the
Luttinger model always vanishes.


The work was supported by NSFC Grant No.10225419.

\appendix

\section{spin conductivity in spin-1/2 representation}

The correlation function of spin conductivity in
Matsubara formalism reads
\begin{equation}
Q_{ij}^{ab} (i\nu)= -\frac{1}{V} \int_{0}^{\beta} du \  e^{i \nu
u} \ave{T_u \hat{J}_i ^{a \dag} (u) \hat{J}_j^b(0)},
\end{equation}
where
\begin{equation}
\hat{J}_i^a (u)=\frac{1}{2}\sum_k C_k^\dag
(u)\Bigl\{\dev{\varepsilon(k)}{k_i}+\dev{d^b}{k_i} \tau^b,\tau^a
\Bigr\} C_k(u).
\end{equation}

After using Wick's theorem and introducing the Matsubara function
$G(k,u)=-\langle T_u C_k(u)C_k^\dag(0) \rangle$, one can obtain
the correlation function
\begin{eqnarray}
&& Q_{ij}^{ab} (i\nu) = \frac{1}{4V\beta}\sum_{k,\omega_n} \nonumber \\
&&  \tr \Bigl(G \Bigl\{\dev{\varepsilon(k)}{k_i}
      +\dev{d^c}{k_i}\tau^c,\tau^a \Bigr\}
  G_+ \Bigl\{\dev{\varepsilon(k)}{k_j}+\dev{d^d}{k_j} \tau^d,\tau^b \Bigr\}\Bigr), \nonumber \\
\end{eqnarray}
where $G$ and $G_+$ refer to $G(k,i\omega)$ and
$G(k,i\omega+i\nu)$, respectively, which can be derived from the
Fourier transform
\begin{equation}
G(k,u)=\frac{1}{\beta}\sum_{\omega_n}G(k,i\omega_n)e^{-i\omega_n
u}.
\end{equation}

In the case
$H_0=\varepsilon(k)+d^a\tau^a$,
\begin{eqnarray}\label{Matsubara fuction}
G(k,i\omega_n) &=& \frac{1}{i\hbar\omega_n+\mu-H_0}
      \nonumber \\
&\equiv& f(k,i\omega_n) (g(k,i\omega_n)+d^a\tau^a ),
\end{eqnarray}
with
\begin{eqnarray}
f(k,i\omega_n) &=& \frac{1}{(i\hbar\omega_n+\mu-\varepsilon)^2 -
|d|^2/4},
      \nonumber \\
g(k,i\omega_n) &=& i\hbar\omega_n+\mu-\varepsilon.
\end{eqnarray}
In the last line of Eq.~(\ref{Matsubara fuction}),
$G(k,i\omega_n)$ is separated into two parts, the U(1) part and
SU(2) part, which facilitates our calculation of the trace term,
\begin{widetext}
\begin{eqnarray}\label{trace in appendixA}
&&
{\tr}\Bigl[\bigl( g(k,i\omega_n)
 +\frac{1}{2} d^c \tau^c \bigr)
   \bigl\{\dev{\varepsilon(k)}{k_i}+\dev{d^c}{k_i}\tau^c,~\tau^a \bigr\}
      \bigl(g(k,i\omega_n+i\nu) + \frac{1}{2}d^c\tau^c\bigr)
       \bigl\{\dev{\varepsilon(k)}{k_j}+\dev{d^d}{k_j} \tau^d,~\tau^b\bigr\}\Bigr]
         \nonumber \\
&=& 2 \Bigl[\Bigl(4
g(k,i\omega_n)
g(k,i\omega_n+i\nu)\delta^{ab}+2d^ad^b-d^2\delta^{ab}
\Bigr)
                \dev{\varepsilon(k)}{k_i}\dev{\varepsilon(k)}{k_j}
            +2i \Bigl(g(k,i\omega_n+i\nu) -g(k,i\omega_n) \Bigr)
                 \epsilon^{abc}d^c \dev{\varepsilon(k)}{k_i}\dev{\varepsilon(k)}{k_j}
      \nonumber \\
&&\hspace{1mm}
+\Bigl(g(k,i\omega_n+i\nu)
+g(k,i\omega_n)\Bigr)
     \Bigl(\dev{\varepsilon(k)}{k_i} \dev{d^b}{k_j}d^a+ \dev{\varepsilon(k)}{k_j} \dev{d^a}{k_i}d^b \Bigr)
     +\Bigl(g(k,i\omega_n) g(k,i\omega_n+i\nu ) + |d|^2/4 \Bigr) \dev{d^a}{k_i}\dev{d^b}{k_j} \Bigr]
\end{eqnarray}

Note that summing the Matsubara function over the frequency gives
\begin{eqnarray}
\frac{1}{\beta}\sum_{\omega_n}
f(k,i\omega_n)f(k,i\omega_n+i\nu)g(k,i\omega_n+i\nu)
    &=&-\frac{1}{\beta} \sum_{\omega_n}
   f(k,i\omega_n)f(k,i\omega_n+i\nu)g(k,i\omega_n)
    \nonumber \\
&=&\frac{i\hbar\nu(n_{F-}-n_{F+})}{|d| \bigl[(i\hbar\nu)^2 - |d|^2
\bigl]},
\end{eqnarray}
and
\begin{eqnarray}
\frac{1}{\beta}
\sum_{\omega_n}
f(k,i\omega_n)f(k,i\omega_n+i\nu)
&=&-\frac{4}{|d|^2\beta} \sum_{\omega_n} f(k,i\omega_n)f(k,i\omega_n+i\nu)g(k,i\omega_n)g(k,i\omega_n+i\nu)
    \nonumber \\
&=&\frac{2(n_{F-}-n_{F+})}{|d| \bigl[(i\hbar\nu)^2 - |d|^2
  \bigl]}.
\end{eqnarray}
Thus the last line of the right-hand side of Eq.~(\ref{trace in
appendixA}) vanishes. Changing $i\nu\rightarrow\omega$, we can
obtain the real and imaginary parts of $Q_{ij}^{ab} (\omega)$,
\begin{eqnarray}
\textrm{Re}\,Q_{ij}^{ab}(\omega)
 = \frac{1}{2V} \!\!\sum_k\!
  \frac{\delta^{ab}|d|^2-d^a d^b}{|d|(\hbar^2\omega^2-|d|^2)}
  \dev{\varepsilon(k)}{k_i} \dev{\varepsilon(k)}{k_j} (n_{F-}
  \!\!-n_{F+}),
            \hspace{17mm}
     \nonumber \\
\textrm{Im}\,Q_{ij}^{ab}
(\omega)
   = \frac{\hbar\omega}{2V} \sum_k\frac{\epsilon^{abc} d^c}{|d|(\hbar^2\omega^2-|d|^2)}
             \dev{\varepsilon(k)}{k_i} \dev{\varepsilon(k)}{k_j} (n_{F-}-n_{F+}).    \hspace{18mm}
\end{eqnarray}
\end{widetext}
Since $ \sigma _{ij} ^{a b} (\omega) =\frac{i}{\hbar\omega} \Bigl[
Q_{ij}^{ab} (\omega) +\frac{\hbar^3 n_0}{4m} \delta^{ab}
\delta_{ij} \Bigr]$, then taking the limit $\omega\rightarrow 0$,
we can obtain the dc conductivity in Eq.~(\ref{eq:spin
conductivity}),
\begin{eqnarray}
\textrm{Re}~\sigma_{ij}^{ab}
 \!=\! -\lim_{\omega\rightarrow 0}\frac{1}{\hbar\omega} \textrm{Im}~Q_{ij}^{ab} (\omega) \hspace{33mm}
      \nonumber \\
 = \!\frac{1}{2V} \sum_k \frac{n_{F-}-n_{F+}}{|d|^3} \epsilon^{abc} d^c
        \dev{\varepsilon(k)}{k_i} \dev{\varepsilon(k)}{k_j}.  \hspace{4mm}
\end{eqnarray}

\section{spin conductivity in spin-3/2 representation}

Before we calculate the spin conductivity in
spin-3/2 representation, it is wise to warm up with the
Clifford algebra. The 4 by 4 gamma matrices
$\Gamma^{\alpha}$ are constructed by the 2 by 2
sigma matrices, which satisfy
$\{\Gamma^\alpha,\Gamma^\beta
\}=2\delta_{\alpha\beta}$
and $\Gamma^1 \Gamma^2
\Gamma^3 \Gamma^4
\Gamma^5=-1$.
Using these gamma matrices, one can also
compose ten antisymmetric matrices
$\Gamma^{\alpha\beta}=\frac{1}{2i}[\Gamma^\alpha,\Gamma^\beta]$.
Together with the identity matrix, $\Gamma^\alpha$ and
$\Gamma^{\alpha\beta}$ span the space of $4\times 4$
hermitian matrices.
The SU(2) generators $\tau^a$ in spin-3/2 representations
can also be expressed as the linear combinations of
$\Gamma^{\alpha\beta}$, \ie,
$\tau^a=\frac{1}{4i}L^a_{\alpha\beta}\Gamma^{\alpha\beta}$
with
\begin{eqnarray}
L^x \!\!&=&\!\!\left(
    \begin{array}{ccccc}
          0&       0 ~&      0~ &      i~&   i\sqrt 3 \\
          0&       0 ~&     -i~ &      0~&   0 \\
          0&       i ~&      0~ &      0~&   0 \\
         -i&       0 ~&      0~ &      0~&   0 \\
  -i\sqrt 3&       0 ~&      0~ &      0~&   0 \\
    \end{array}
    \right),
    \nonumber\\[2mm]
L^y \!\!&=&\!\!\left(
    \begin{array}{ccccc}
          0&       0 &      ~i~~ &      0&    0\\
          0&       0 &      ~0~~ &      i&   -i\sqrt 3 \\
         -i&       0 &      ~0~~ &      0&    0 \\
          0&      -i &      ~0~~ &      0&    0 \\
          0& i\sqrt 3&      ~0~~ &      0&    0 \\
    \end{array}
    \right),
    \nonumber\\[2mm]
L^z \!\!&=&\!\!\left(
    \begin{array}{ccccc}
          ~~0~&       -i&      ~0&      0&    ~0~~\\
          ~~i~&       0 &      ~0&      0&    ~0~~ \\
          ~~0~&       0 &      ~0&    -2i&    ~0 ~~\\
          ~~0~&       0 &     ~2i&      0&    ~0~~ \\
          ~~0~&       0 &      ~0&      0&    ~0~~ \\
    \end{array}
    \right).
\end{eqnarray}
Note that $L^a$ are antisymmetric and satisfy the commutation
relation $[L^a,L^b]=i\epsilon^{abc}L^c$. Thus they form the spin-2
representation of the SU(2) algebra. The following formulas are
inevitable in further calculations,
\begin{eqnarray}
&&\tr(\Gamma^\alpha \Gamma^\beta) = 4\delta^{\alpha\beta},
 \hspace{16mm}
   \tr(\Gamma^\alpha \Gamma^\beta \Gamma^\gamma) = 0,
     \nonumber \\
&&\tr(\Gamma^\alpha \Gamma^\beta \Gamma^\mu\Gamma^\nu)
    = 4(\delta^{\alpha\beta}\delta^{\mu\nu}\!
   -\delta^{\alpha\mu}\delta^{\beta\nu}\!
    +\delta^{\alpha\nu}\delta^{\beta\mu}),
       \nonumber \\
&&\tr(\Gamma^\alpha \Gamma^\beta \Gamma^\gamma
\Gamma^\mu \Gamma^\nu) = -4\epsilon^{\alpha\beta\gamma\mu\nu}\!\!,
      \hspace{2mm}
   \tr(\Gamma^{\alpha\beta} \Gamma^\gamma) = 0,
      \nonumber \\
&&\tr(\Gamma^{\alpha\beta}\Gamma^\mu \Gamma^\nu)
      = 4i(\delta^{\alpha\mu}\delta^{\beta\nu}
       -\delta^{\alpha\nu}\delta^{\beta\mu}),
            \nonumber \\
&&\tr(\Gamma^{\alpha\beta} \Gamma^\gamma \Gamma^\mu
 \Gamma^\nu) = 4i\epsilon^{\alpha\beta\gamma\mu\nu}.
\end{eqnarray}

We take the unperturbed and perturbed parts of the Hamiltonian to
be $H_0=\sum_k
C^\dag_k\bigl(\varepsilon(k)+d^\alpha\Gamma^\alpha\bigr)C_k$ and
$H'=\sum_k C^\dag_k\bigl(h^\beta\Gamma^\beta\bigr)C_k$.
Accordingly, the Kubo formula for the spin conductivity reads
\begin{equation}
\ave{J^b_i(\mathbf r,t)}=\sigma^{b\alpha}_i(\mathbf q,\omega)
h^\alpha(\mathbf r,t).
\end{equation}
In the limit $q\rightarrow 0$,
\begin{equation}
\sigma _i ^{b\alpha}
(\omega) =
 \frac{1}{\hbar\omega V} \int _{-\infty}^{~t}\!\!dt' \  e^{i \omega (t-t')}
  \ave{[\hat{J}_i^b (t), \hat\Gamma^\alpha (t') ] }_0 ,
\end{equation}
and the corresponding retarded correlation function in Matsubara
formalism is given by
\begin{equation}
 Q_i^{b\alpha} (\omega)= -\frac{1}{V} \int_{0}^{\beta} du
 e^{i \nu u} \ave{T_u \hat{J}^{b}_i(u) \hat\Gamma^\alpha (0)}_0.
\end{equation}
Similarly, introducing the
Matsubara function
$G(k,i\omega_n)$ and using
the definition of spin
current
\begin{equation}
\hat{J}_i ^a = \frac{1}{2}
  \sum_k C_k^\dag \Bigl\{\dev{\varepsilon(k)}{k_i}+\dev{d^\beta}{k_i}\Gamma^\beta,\tau^a \Bigr\} C_k,
\end{equation}
we can calculate the trace
term
\begin{widetext}
\begin{eqnarray}\label{trace in appendixB}
&&
\frac{1}{4i}\tr\Bigl[\bigl(g(k,i\omega_n)
+d^\mu \tau^\mu \bigr)
     \bigl\{\dev{\varepsilon(k)}{k_i}+\dev{d^\mu}{k_i} \tau^\mu,~
      L^b_{\beta\gamma}\Gamma^{\beta\gamma} \bigr\}
      \bigl(g(k,i\omega_n+i\nu) +d^\mu \tau^\mu \bigr) \Gamma^\alpha
             \Bigr]
    \nonumber \\
&=&
2\Bigl(g(k,i\omega_n+i\nu)-g(k,i\omega_n)
\Bigr)
       \Bigl(2\dev{\varepsilon(k)}{k_i}L^a_{\alpha\beta}d^\beta
              -\epsilon^{\alpha\beta\gamma\mu\nu}L^a_{\beta\gamma}\dev{d^\mu}{k_i}d^\nu \Bigr).
\end{eqnarray}

Note that $L^a$ are all imaginary, then the dc conductivity is given by
\begin{equation}
\sigma^{b\alpha}_i =
 \frac{-1}{4V}\sum_k
  \textrm{Im}\Bigl(2\dev{\varepsilon}{k_i}L^b_{\alpha\beta}d^\beta
  -\epsilon^{\alpha\beta\gamma\mu\nu}L^b_{\beta\gamma}\dev{d^\mu}{k_i}d^\nu \Bigr)
  \frac{n_{F_-}-n_{F_+}}{|d|^3}.
\end{equation}
\end{widetext}

\end{document}